\documentclass[twocolumn,tighten]{aastex62}
\usepackage{booktabs}
\usepackage{amsmath}

\newcommand{\keyw}[1]{\textcolor{gray}{#1}}

\newcommand{\emcee}{\texttt{emcee}}
\newcommand{\CASA}{\texttt{CASA}}

\begin{document}

% figures

% fig 1: Dust emission map, and CO moment-zero map.

% fig 2: Deprojected radial profile of dust and CO

% (fig: map in polar coordnates to show how axysymmetric the disk is?)

% fig 3. Visibility plot: obs in black with best-fit model overlaid in red.

% fig: overlay of CO on dust emission map

% fig 4: Best fit model: Model (left) and residuals (right) 

\title{Disk Substructures at High Angular Resolution Program (DSHARP):\\
  VIII. The Rich Ringed Substructures in the AS 209 Disk }  % S.A
%  X. Complex multi-ring structure in the AS 209 disk } % JH
%  X. Six nested rings in 'Russian Doll' disk AS 209 } % KIO

\author{Viviana V.~Guzm{\'a}n}
\affiliation{Joint ALMA Observatory, Avenida Alonso de C{\'o}rdova 3107, Vitacura, Santiago, Chile}
\affiliation{Instituto de Astrof{\'i}sica, Ponticia Universidad Cat{\'o}lica de Chile, Av.~Vicu{\~n}a Mackenna 4860, 7820436 Macul, Santiago, Chile}

\author{Jane Huang}
\affiliation{Harvard-Smithsonian Center for Astrophysics, 60 Garden Street, Cambridge, MA 02138, USA}

\author{Sean~M.~Andrews}
\affiliation{Harvard-Smithsonian Center for Astrophysics, 60 Garden Street, Cambridge, MA 02138, USA}

\author{Andrea Isella}
\affiliation{Department of Physics and Astronomy, Rice University, 6100 Main Street, Houston, TX 77005, USA}

\author{Laura M.~P{\'e}rez}
\affiliation{Departamento de Astronom{\'{\i}}a, Universidad de Chile, Casilla 36-D, Santiago, Chile}

\author{John M.~Carpenter}
\affiliation{Joint ALMA Observatory, Avenida Alonso de C{\'o}rdova 3107, Vitacura, Santiago, Chile}

%\author{Nicol{\'a}s T.~Kurtovic}
%\affiliation{Departamento de Astronom{\'{\i}}a, Universidad de Chile, Casilla 36-D, Santiago, Chile}

\author{Cornelis P.~Dullemond}
\affiliation{Zentrum f{\"u}r Astronomie, Heidelberg University, Albert Ueberle Str.~2, 69120 Heidelberg, Germany}

%\author{A.~Meredith Hughes}
%\affiliation{Department of Astronomy, Van Vleck Observatory, Wesleyan University, 96 Foss Hill Drive, Middletown, CT 06459, USA}

\author{Luca Ricci}
\affiliation{Department of Physics and Astronomy, California State University Northridge, 18111 Nordhoff Street, Northridge, CA 91130, USA}

%\author{Erik Weaver}
%\affiliation{Department of Physics and Astronomy, Rice University, 6100 Main Street, Houston, TX 77005, USA}

\author{Tilman Birnstiel}
\affiliation{University Observatory, Faculty of Physics, Ludwig-Maximilians-Universit\"at M\"unchen, Scheinerstr.~1, 81679 Munich, Germany}

\author{Shangjia Zhang}
\affiliation{Department of Physics and Astronomy, University of Nevada, Las Vegas, 4505 S.~Maryland Pkwy, Las Vegas, NV 89154, USA}

\author{Zhaohuan Zhu}
\affiliation{Department of Physics and Astronomy, University of Nevada, Las Vegas, 4505 S.~Maryland Pkwy, Las Vegas, NV 89154, USA}

\author{Xue-Ning Bai}
\affiliation{Institute for Advanced Study and Tsinghua Center for Astrophysics, Tsinghua University, Beijing 100084, China}

\author{Myriam Benisty}
\affiliation{Unidad Mixta Internacional Franco-Chilena de Astronom\'{i}a, CNRS/INSU UMI 3386, Departamento de Astronom{\'i}a, Universidad de Chile, Camino El Observatorio 1515, Las Condes, Santiago, Chile}
\affiliation{Univ.~Grenoble Alpes, CNRS, IPAG, 38000 Grenoble, France}

\author{Karin I.~{\"{O}}berg}
\affiliation{Harvard-Smithsonian Center for Astrophysics, 60 Garden Street, Cambridge, MA 02138, USA}

\author{David J.~Wilner}
\affiliation{Harvard-Smithsonian Center for Astrophysics, 60 Garden Street, Cambridge, MA 02138, USA}

\begin{abstract}
We present a detailed analysis of the high-angular resolution
(0$\farcs$037, corresponding to 5~au) observations of the 1.25~mm
continuum and $^{12}$CO $2-1$ emission from the disk around the
T~Tauri star AS~209. AS~209 hosts one of the most unusual disks from
the DSHARP sample, the first high angular resolution ALMA survey of
disks \citep{Andrews2018}, as nearly all of the emission can be
explained with concentric Gaussian rings. In particular, the dust
emission consists of a series of narrow and closely spaced rings in
the inner $\sim60$~au, two well-separated bright rings in the outer
disk, centered at 74 and 120~au, and at least two fainter emission
features at 90 and 130~au. We model the visibilities with a parametric
representation of the radial surface brightness profile, consisting of
a central core and 7 concentric Gaussian rings. Recent hydro-dynamical
simulations of low viscosity disks show that super-Earth planets can
produce the multiple gaps seen in AS~209 millimeter continuum
emission. The $^{12}$CO line emission is centrally peaked and extends
out to $\sim300$~au, much farther than the millimeter dust
emission. We find axisymmetric, localized deficits of CO emission
around four distinct radii, near 45, 75, 120 and 210~au. The outermost
gap is located well beyond the edge of the millimeter dust emission,
and therefore cannot be due to dust opacity and must be caused by a
genuine CO surface density reduction, due either to chemical effects
or depletion of the overall gas content.
\end{abstract}

\keywords{\keyw{circumstellar matter --- planetary systems: formation, protoplanetary disks --- dust}}

%%%%%%% figures

\newcommand{\FigImCont}{%
\begin{figure}[t!]
  \centering
  \includegraphics[width=0.5\textwidth]{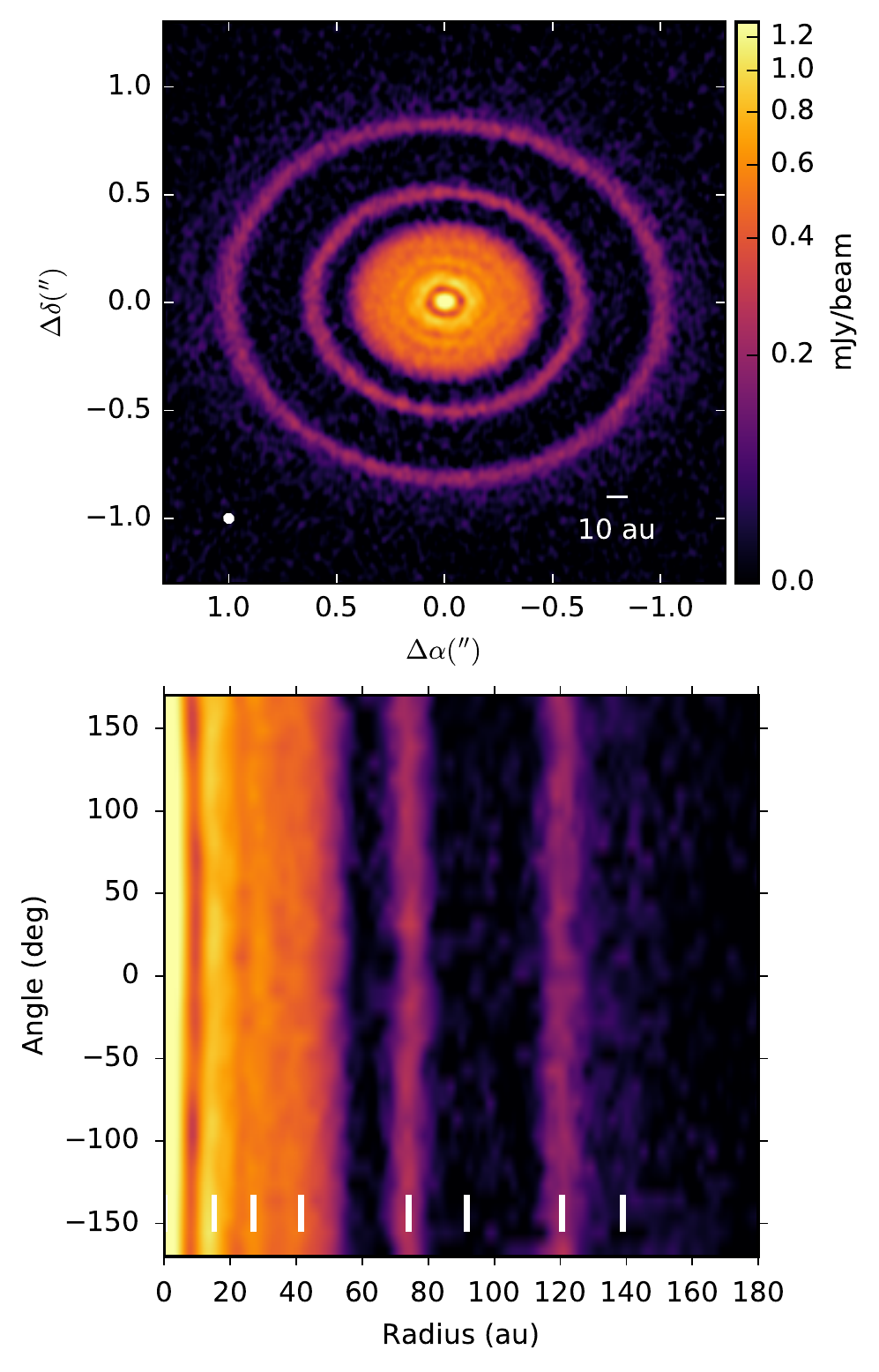}
  \caption{The dust continuum emission map (upper panel), and the
    deprojected emission shown in polar coordinates (bottom panel),
    using an inclination of $34.88^\circ$ and a position angle of
    $85.76^\circ$. The beam of $0\farcs03 \times 0\farcs04$ is shown in
    the bottom left, corresponding to a spatial resolution of $3.6
    \times 4.8$~au. The white vertical bars in the bottom panel locate
    the position of the rings.}
  \label{fig:im}
\end{figure}
}

\newcommand{\FigCO}{%
\begin{figure}[h!]
  \centering
  \includegraphics[width=0.5\textwidth]{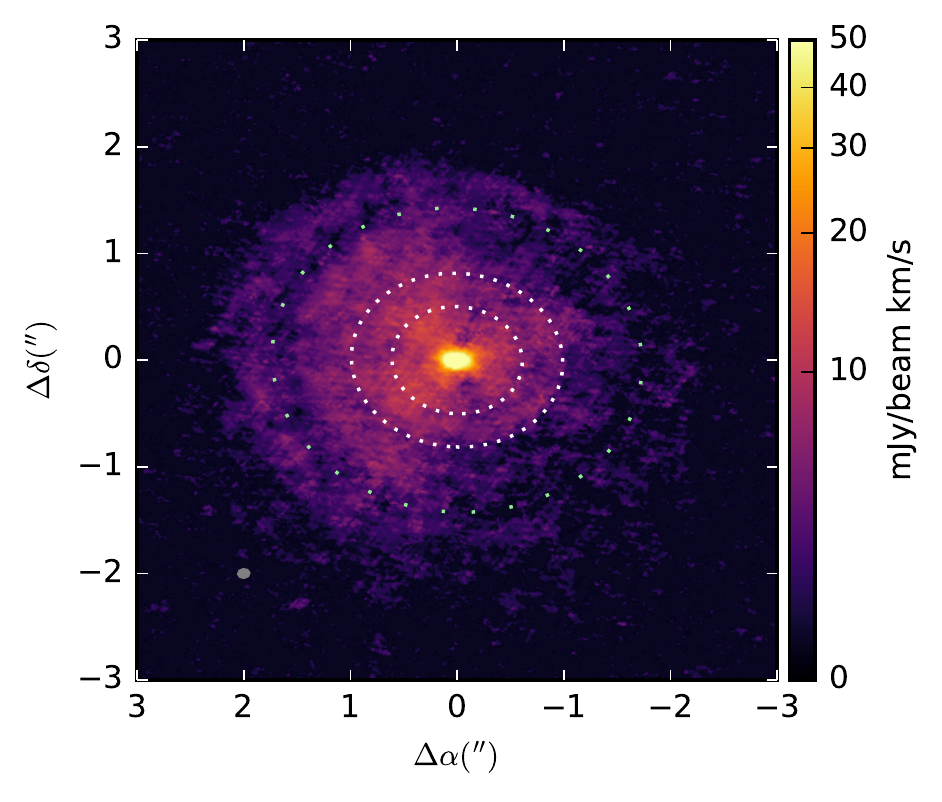}
  \includegraphics[width=0.5\textwidth]{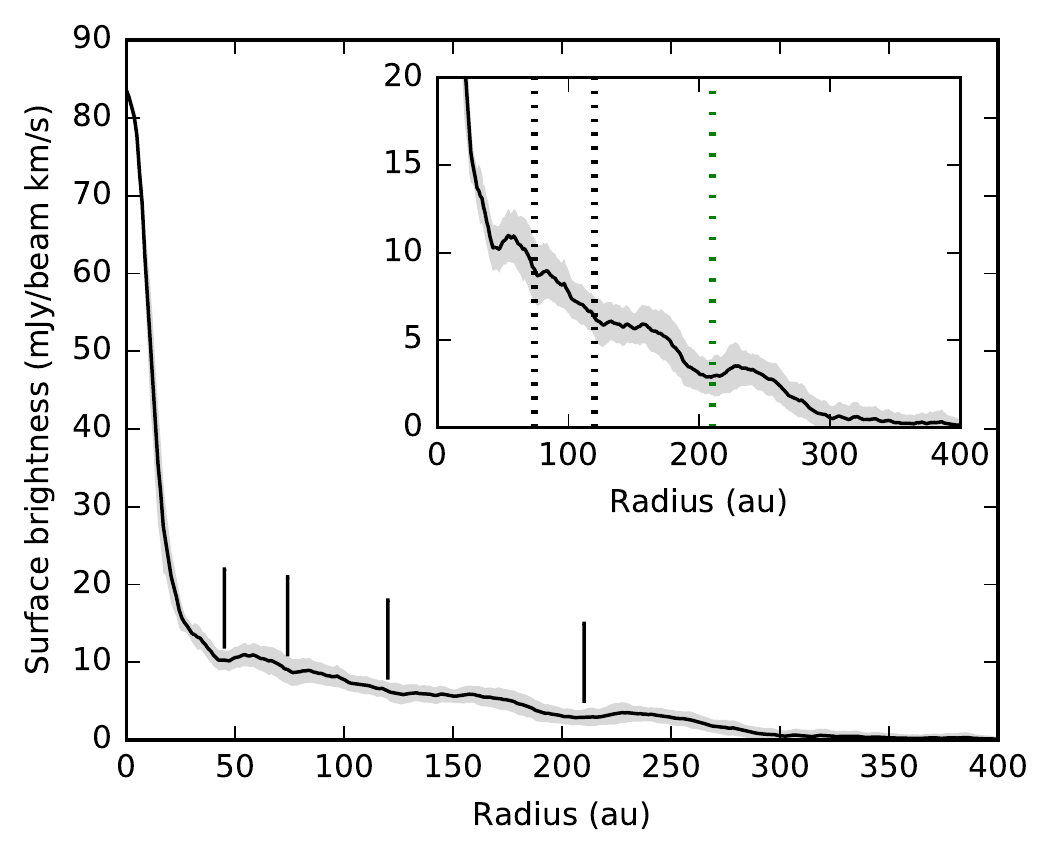}
  \caption{Moment-zero map (upper) and azimuthally-averaged
    deprojected radial profile (bottom) of the $^{12}$CO $2-1$ line
    emission. Emission with signal-to-noise ratio lower than 3 has
    been clipped. Only the uncontaminated East part of the image was
    considered to create the deprojected profile. The four vertical
    bars mark the position of the CO gaps. The two white (black in
    lower panel) dotted lines mark the position of the prominent outer
    dust rings, located at 75 and 120~au, and the green dotted line
    marks the position of the outermost CO gap near 210~au.}
  \label{fig:co-im}
\end{figure}
}

\newcommand{\FigCOpanels}{%
\begin{figure}[h!]
  \centering
  \includegraphics[width=0.5\textwidth]{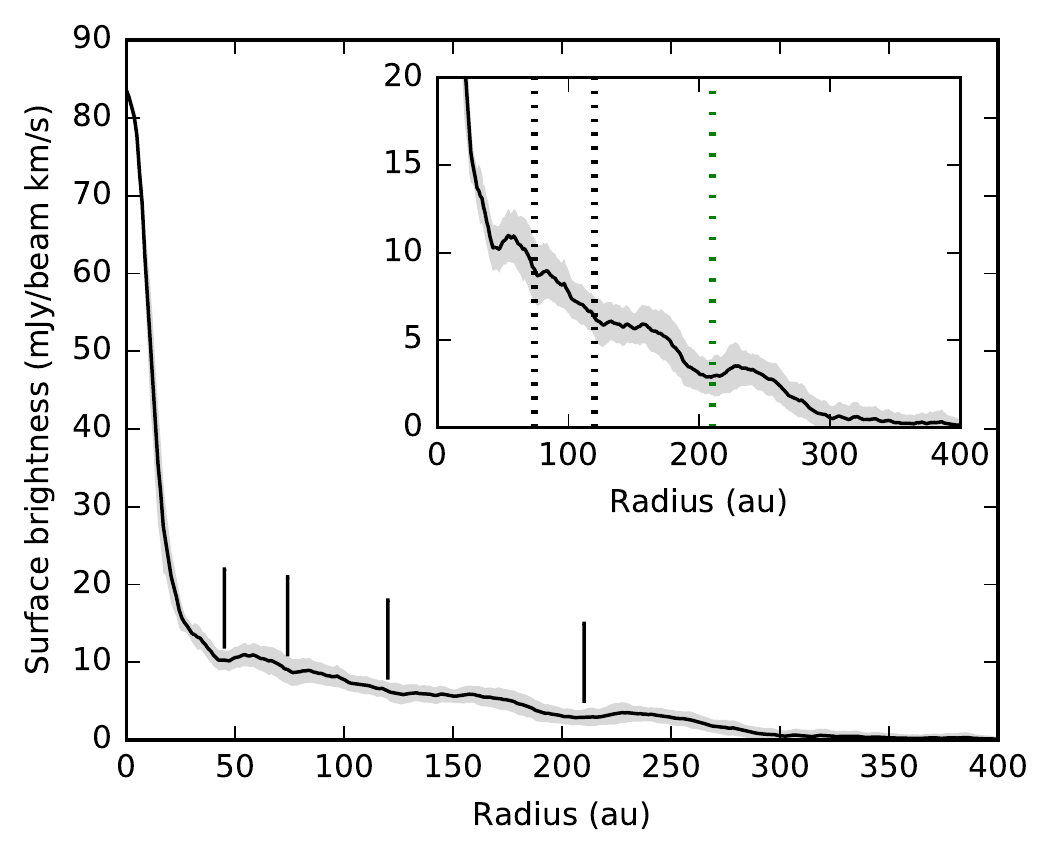}
  \caption{Top: Moment-zero map of the $^{12}$CO $2-1$ line
    emission. The white dotted lines mark the position of the two
    brighter outer dust rings. Middle: CO map shown in polar
    coordinates. Bottom: Deprojected radial profile of the CO
    emission. Only the East side of the disk (angles from 0 to 150
    degrees) is considered to obtain the radial profile.}
  \label{fig:co-panels}
\end{figure}
}

\newcommand{\FigCOchans}{%
\begin{figure*}[t!]
  \centering
  \includegraphics[width=\textwidth]{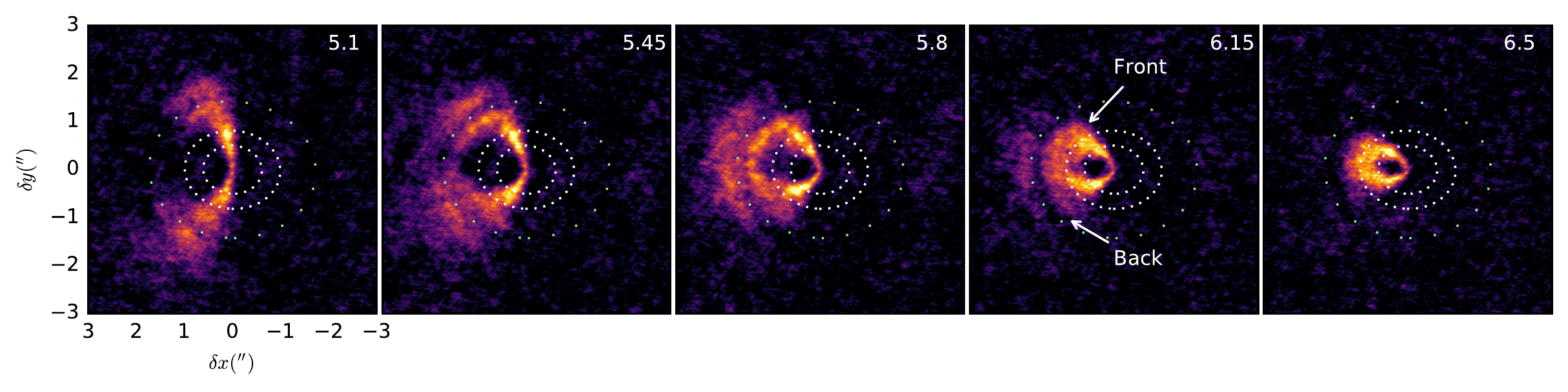}
  \caption{Channel-maps of the $^{12}$CO $2-1$ line emission. Only
    channels from the uncontaminated East side of the disk are
    shown. The velocity of each channel is shown in top right
    corner. The two inner ellipses (dotted lines) mark the position of
    the two brighter outer dust rings located at 74 and 120~au. A
    third ellipse (green dotted line) corresponding to a projected
    radius of 210~au is also drawn, to mark the position of a gap of
    CO emission in the outer disk.}
  \label{fig:co-chans}
\end{figure*}
}

\newcommand{\FigCOchansALL}{%
\begin{figure*}[h!]
  \centering
  \includegraphics[width=\textwidth]{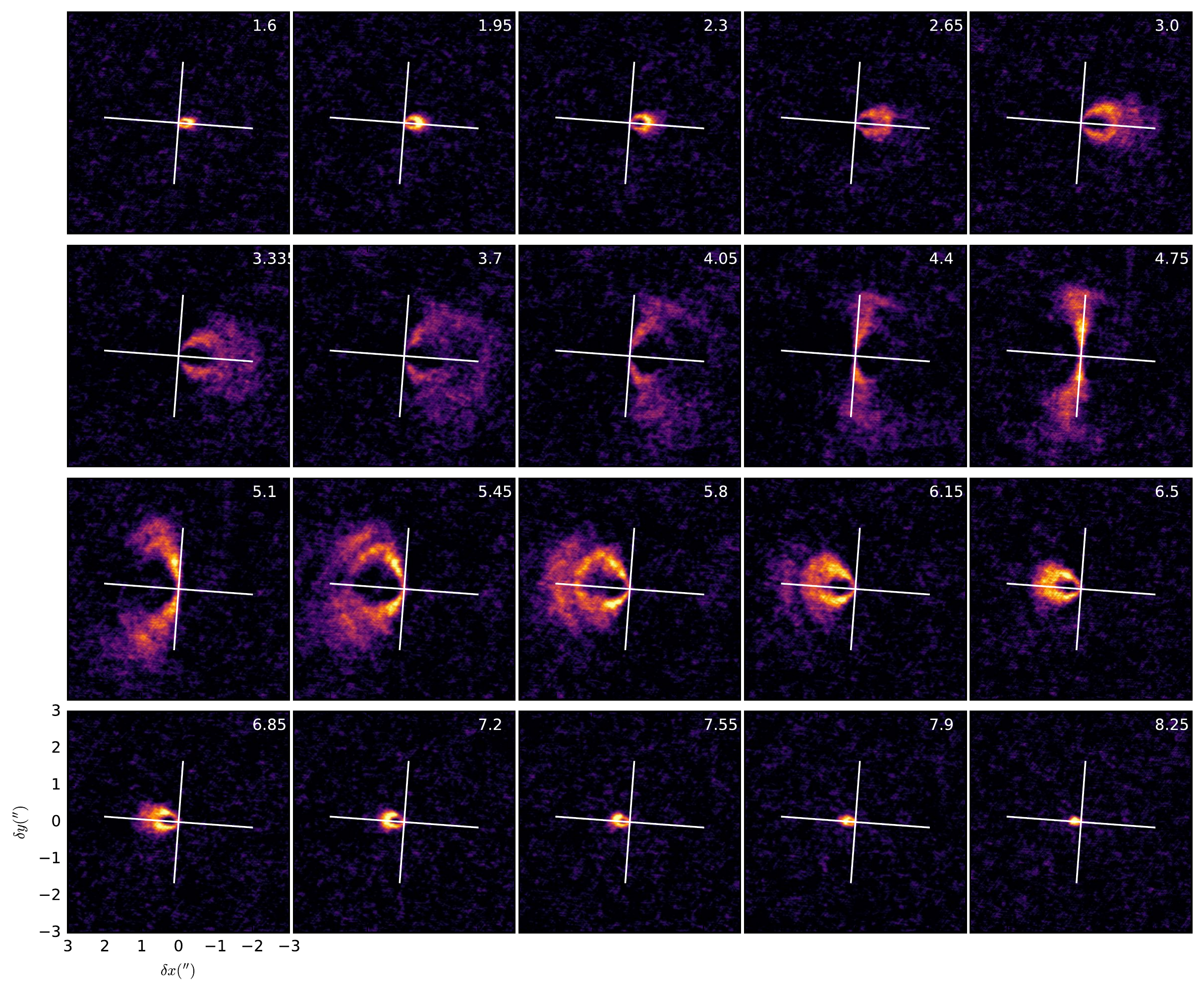}
  \caption{Channel-maps of the $^{12}$CO $2-1$ line emission. The
    velocity of each channel is shown in top right corner. The major
    and minor axis are shown in white.}
  \label{fig:co-chans-all}
\end{figure*}
}

\newcommand{\FigPRofile}{%
\begin{figure}[t!]
  \centering
  \includegraphics[width=0.5\textwidth]{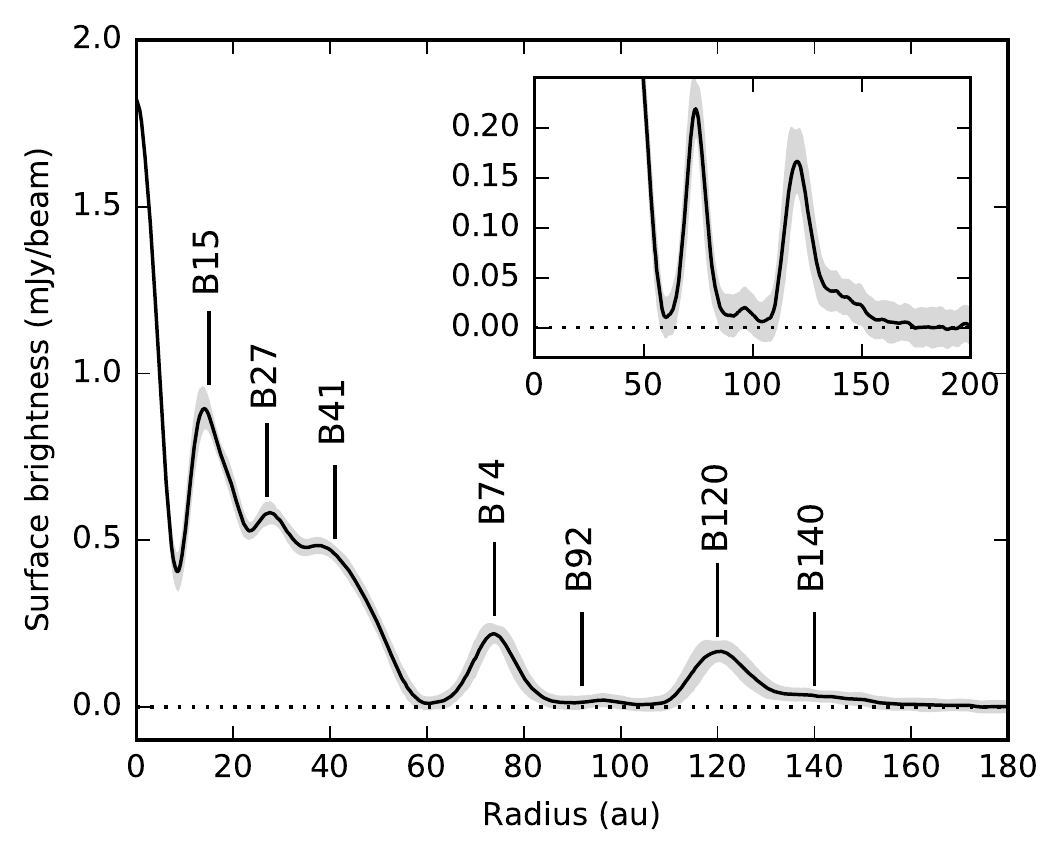}
  \caption{Deprojected azimuthally-averaged radial profile of the dust
    continuum emission. An inset is shown in the upper right panel to
    see the faint emission. The gray ribbon shows the standard
    deviation at each radius, and the vertical bars show the location
    of the rings given by our best-fit model.}
  \label{fig:rad-prof}
\end{figure}
}

\newcommand{\FigUVprof}{%
\begin{figure}[b!]
  \centering
  \includegraphics[width=0.5\textwidth]{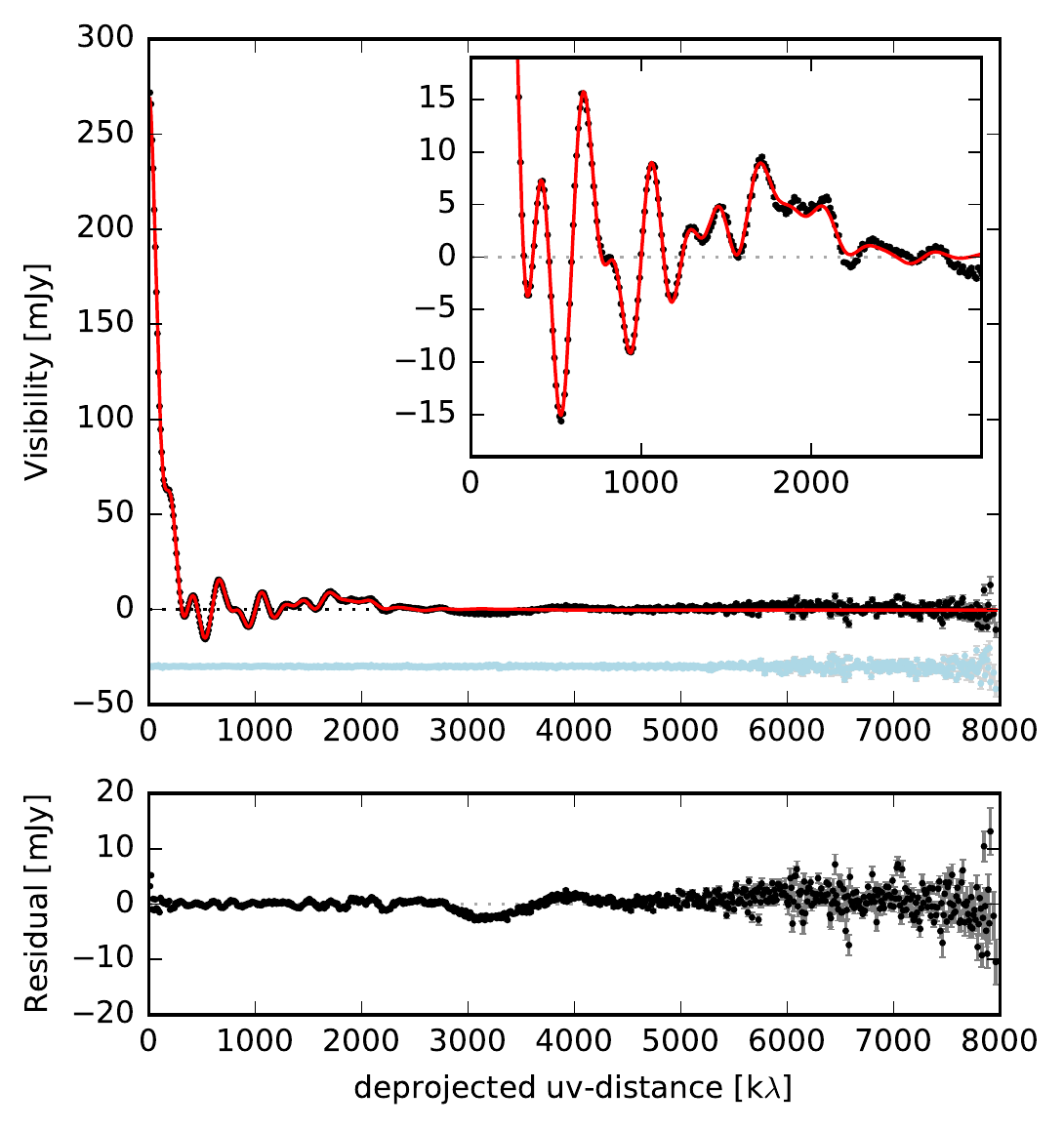}
  \caption{Deprojected and radially averaged visibilities. The real
    part of the visibilities are shown in black and the imaginary part
    are shown in light-blue (shifted to $-30$~mJy). The best-fit model
    is shown in red. An inset is shown in the upper right corner.}
  \label{fig:uv-prof}
\end{figure}
}

\newcommand{\FigModel}{%
\begin{figure*}
  \centering
  \includegraphics[width=\textwidth]{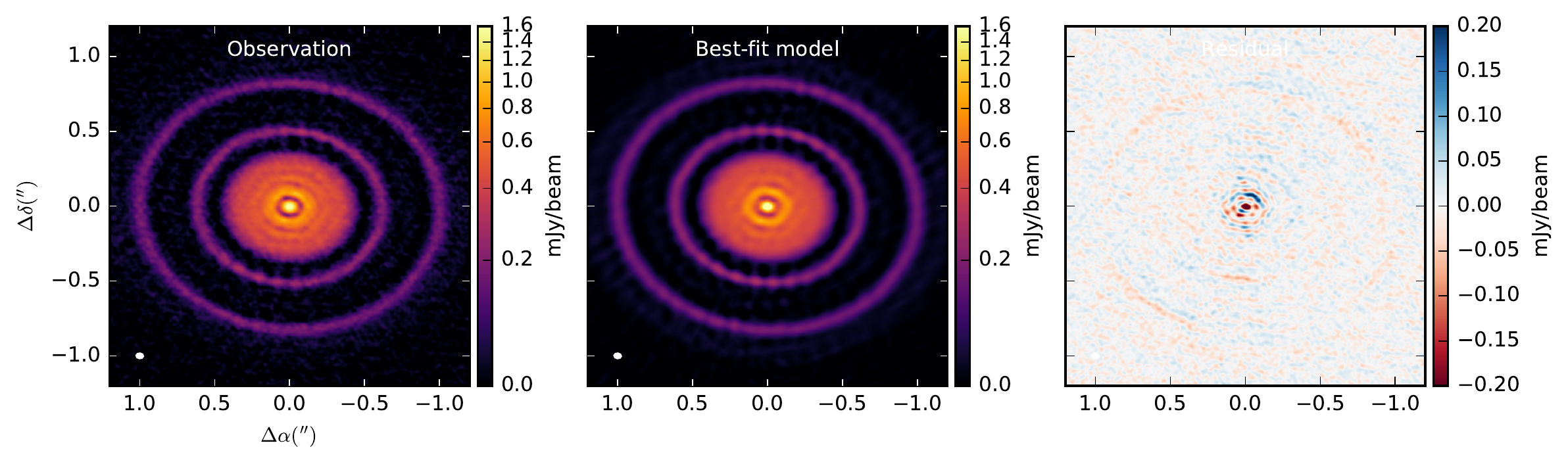}
  \caption{Observed dust continuum emission (left), best-fit model
    (middle) and residual (right) maps. The residuals are shown in
    linear scale}
  \label{fig:model-im}
\end{figure*}
}

\newcommand{\FigSimulation}{%
\begin{figure}[b!]
  \centering
  \includegraphics[width=0.5\textwidth]{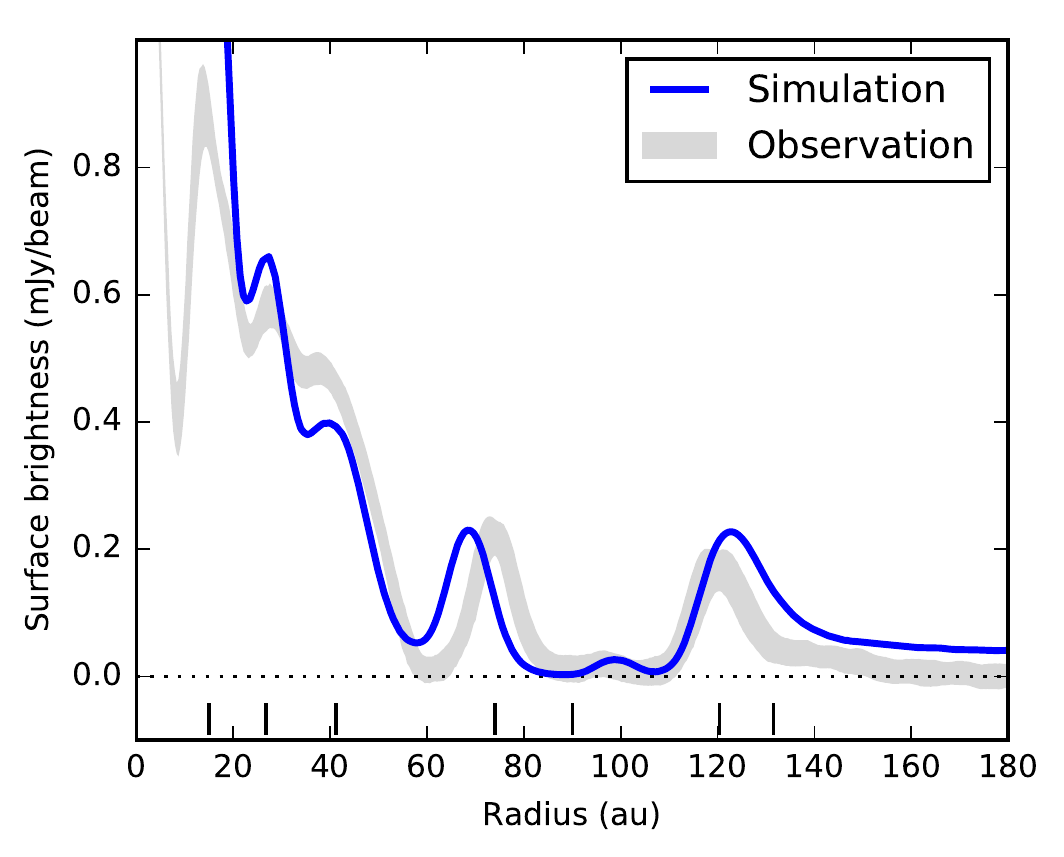}
  \caption{Hydrodynamical simulation showing the gaps induced in a
    disk by the presence of a planet orbiting at 99~au from the star,
    overlaid on the observed radial profile in AS~209. A detailed
    description of the simulation is given in \citet{Zhang2018}.}
  \label{fig:simulation}
\end{figure}
}

\newcommand{\TabBestFitParamsT}{%
   \begin{table}%[b!]
     \begin{center}
       \caption{Best-fit parameters.} 
       \label{tab:bf}
       \begin{tabular}{ccrr}\toprule
         \multicolumn{4}{c}{\it Disk geometry} \\
         $i$ (deg)  & PA (deg) & $\delta_x$ (mas) & $\delta_y$ (mas) \\
         $34.883_{-0.0042}^{+0.0151}$ & $85.764_{-0.0061}^{+0.0184}$ & $1.699_{-0.0006}^{+0.0003}$ & $-3.102_{-0.0003}^{+0.0006}$ \\
         \midrule
         \multicolumn{4}{c}{\it Gaussian components} \\
         Ring  Name$^a$  & Rel. Amp$^b$  & $r_i$ (au) & FWHM (au)  \\
         B0        & $1.000_{-0.0027}^{+0.0024}$ & $0.00$ & $6.69_{-0.01}^{+0.01}$ \\
         B15  (B14)  & $0.274_{-0.0012}^{+0.0005}$ & $15.13_{-0.03}^{+0.02}$ & $7.41_{-0.01}^{+0.03}$ \\
         B27  (B28)  & $0.133_{-0.0012}^{+0.0014}$ & $27.07_{-0.03}^{+0.02}$ & $11.72_{-0.03}^{+0.05}$ \\
         B41  (B38)  & $0.115_{-0.0001}^{+0.0001}$ & $41.42_{-0.03}^{+0.05}$ & $17.40_{-0.06}^{+0.03}$ \\
         B74  (B74)  & $0.074_{-0.0002}^{+0.0003}$ & $74.08_{-0.01}^{+0.02}$ & $7.34_{-0.01}^{+0.01}$ \\
         B92  (B99)  & $0.004_{-0.0001}^{+0.0001}$ & $91.76_{-0.09}^{+0.14}$ & $23.39_{-0.08}^{+0.06}$ \\
         B120 (B120) & $0.051_{-0.0002}^{+0.0003}$ & $120.42_{-0.02}^{+0.02}$ & $9.84_{-0.03}^{+0.02}$ \\
         B140 (B141) & $0.008_{-0.0001}^{+0.0001}$ & $139.06_{-0.08}^{+0.07}$ & $23.10_{-0.16}^{+0.08}$ \\
         \bottomrule
       \end{tabular}\\
     \end{center}
       $^a$ The name in parenthesis corresponds to the one adopted in
     \citet{Huang2018a}.\\
     $^b$ Amplitudes are normalized to $A_0$, so that
       $I_{norm}(r) = I(r)/A_0$. The normalization factor is given by
       $A_0 = F_{tot} / ( 2 \pi \int^\infty_0 I_{norm}(r) J_0(0) r dr
       ) $, where $F_{tot}$ is the disk integrated flux density ($F_{tot}
       \equiv V(\rho=0) = 0.27$ Jy.)

     \label{tab:best-fit}
   \end{table}
 }

%%%%%%%

\section{Introduction}
\label{sec:intro}

The distribution of gas and dust in protoplanetary disks will directly
impact the outcome of planetary systems
\citep{Weidenschilling1977,Oberg2011}. Characterizing the spatial
distribution of both the dust and gas in disks is therefore essential
to understanding how and what kind of planets can form. The main
problem theoretical models currently have is the fast migration of
mm-sized dust particles towards the central star, preventing the
formation of planetesimals, especially at larger distances from the
star \citep{Takeuchi2002,Takeuchi2005,Brauer2007,Brauer2008}. The
solution that has been invoked to solve this problem is the presence
of local pressure maxima that can stop, at least temporarily, the
migration of solid particles and concentrate them for enough time to
allow them to grow and form larger bodies
\citep[e.g.,][]{Pinilla2012}. High-angular resolution ALMA
observations of the millimeter dust continuum emission have shown
evidence of such substructures in a handful of disks around nearby
young stars, including HL~Tau \citep{ALMAPart2015}, HD~163296
\citep{Isella2016}, TW~Hya \citep{Andrews2016}, Elias~24
\citep{Cieza2017,Dipierro2018} and AS~209 \citep{Fedele2018}. All of
these disks present multiple ring/gap structures, and demonstrate the
presence of mm-sized grains out to radii of at least 100~au in the
disks. The origin of these ring-like substructures is unclear.  The
most favored mechanisms include planet-disk interactions
\citep[e.g.][]{Dong2017}, radial pressure variations due to zonal
flows in MHD turbulent disks \citep{Johansen2009}, and
snowline-induced gaps \citep{Zhang2015}.

While the presence of substructures seems common, a larger sample of
disks is needed to fully characterize the prevalence and configuration
of such substructures, and constrain the physical or chemical processes
responsible for them. This is the motivation of the Disk Substructures
at High Angular Resolution Project (DSHARP), one of the large programs
carried out with the Atacama Large Millimeter Array (ALMA) in Cycle
4. The goal of the project is to characterize in an homogeneous way
the substructures of 20 nearby protoplanetary disks, by mapping the
240~GHz dust continuum emission at a resolution of 35~mas,
corresponding to 5~au \citep{Andrews2018}. One of the main outcomes of
this survey is that bright rings and relatively faint gaps are an
extremely common feature of disks, but the configuration (position and
contrast) of the rings varies substantially from source to source
\citep{Huang2018a}.

\FigImCont{}

In this paper we focus on one of the most unusual DSHARP sources, the
disk around the classical T~Tauri star AS~209. The large number of
rings, the narrowness of the rings, and the wide gaps in the outer
disk make AS~209 especially intricate compared to the vast majority of
disks observed at high-angular resolution. The star is located to the
northeast of the main Ophiuchus star-forming region, at a distance of
$121\pm2$~pc \citep{Gaia2018}. The star has a spectral type K5, a mass
of 0.9~M$_\odot$, and an age of 1.6~Myr \citep[see Table~1
  in][]{Andrews2018}. Observations at different wavelengths clearly
show the effect of radial drift of larger grains, as the emission is
noticeably more compact at longer wavelengths
\citep{Perez2012,Tazzari2016}. The surface density profile has been
characterized with 870~$\mu$m observations at 0\farcs3 angular
resolution \citep{Andrews2009}. More recently, \citet{Fedele2018}
presented ALMA observations of the disk at $\sim$0\farcs17 angular
resolution. The emission was characterized by a bright central
component and the presence of two weaker dust rings near 75 and
130~au, and two gaps near 62 and 103~au. No obvious substructure was
observed in the inner 60~au disk, except for a kink around
$20-30$~au. \citet{Fedele2018} also presented hydro-dynamical
simulations and found that the gap near 100~au located between the two
outer rings could be produced by a Saturn-like planet.

%% Observations of the scattered light emission from the AS~209 disk have
%% been recently presented by \citet{Avenhaus2018}. The emission is
%% surprisingly faint and compact compared to other disks of similar dust
%% mass and age. Moreover, AS~209 presents a large IR excess, which
%% usually results in bright scattered light emission. This could be the
%% result of self-shadowing of the inner disk, although the most likely
%% alternative is that the disk is very settled which hampers the
%% efficient scattering of photons from the star.

The disk has also been observed in molecular line
emission. \citet{Huang2016} presented observations of the three main
CO isotopologues at 0\farcs6 angular resolution. While the emission
from the most abundant isotopologue, $^{12}$CO, was found to be
centrally peaked and decreasing monotonically with radius, the
$^{13}$CO and C$^{18}$O emission showed evidence of an outer ring or
bump centered at 150~au, near the millimeter dust edge.

%Other molecules have also been
%detected, including DCN, DCO$^+$, HCO$^+$ \citep{Huang2017}, HCN
%isotopologues \citep{Guzman2017},

In this work we present 1.25~mm dust continuum and $^{12}$CO $2-1$
line observations in the AS~209 disk. The observations and data
reduction are presented in section \ref{sec:obs}. The results of the
dust continuum emission and the $^{12}$CO line emission are described
in section \ref{sec:results}. A discussion is presented in
section~\ref{sec:discussion} and a summary is given in
section~\ref{sec:summary}.

\section{Observations}
\label{sec:obs}

The observations presented here are part of the DSHARP ALMA Large
Program (2016.1.00484.L). The AS~209 disk was observed
with ALMA in Band~6 in September 2017 in configurations
C40-8/9. Shorter-baselines were observed in May 2017 in configuration
C40-5. We use additional archival data from projects 2013.1.00226
\citep{Huang2016} and 2015.1.00486.S \citep{Fedele2018}, that provide
information on shorter and intermediate baselines, respectively.

A detailed description of the observations and data reduction process
can be found in \citet{Andrews2018}. Briefly, we first self-calibrated
the short baselines, and in a second step self-calibrated the combined
observations of short- and long-baselines. The continuum observations
were cleaned in \CASA{}~5.1, using the multi-scale option and a robust
parameter of $-0.5$. A taper with FWHM of 37~mas~$\times$~10~mas and
position angle (PA) of 162$^{\circ}$ was used to minimize PSF-related
artifacts. The resulting continuum image has a beam of
38~mas~$\times$~36~mas, PA of $68^{\circ}$ and a rms noise of
19~$\mu$Jy~beam$^{-1}$.

The solutions of the continuum self-calibration were then applied to
the $^{12}$CO data. The molecular line data was first regridded to
channels of 0.35~km~s$^{-1}$ and then cleaned with a robust parameter
of 1.0. A Keplerian mask was used to help the cleaning process. The
resulting $^{12}$CO image has a beam of 95~mas~$\times$~70~mas, PA of
$97^{\circ}$ and a rms noise of 0.8~mJy~beam$^{-1}$ per channel.

\section{Results}
\label{sec:results}

In this section we first describe the main features of the dust
continuum emission and model the surface brightness in the {\it uv}-plane to
extract the positions, widths and amplitudes of the different ring
components. We then describe the spatial distribution of the $^{12}$CO
emission and its relation to the dust continuum emission.

\subsection{Dust continuum emission}

\FigPRofile{}

The image of the 1.25~mm dust continuum emission from the AS~209 disk
is shown in Figure~\ref{fig:im}. The map is shown in the upper panel
while the lower panel shows the emission in polar coordinates to
better visualize the axisymmetric nature of the emission. To create
the map in polar coordinates, the image is first deprojected using an
inclination of $34.88^\circ$ and a position angle of $85.76^\circ$
(see section~3.1.1). The polar angle increases in the clockwise
direction, where zero degrees corresponds to the minor axis of the
disk in the south direction. The dust emission is characterized {\em
  almost entirely} by a series of concentric narrow rings and
gaps. Although the emission looks very axisymmetric, we cannot exclude
the possibility that the rings have a small eccentricity
\citep[$e<0.15$; see discussion
  in][]{Huang2018a,Zhang2018}. Fig.~\ref{fig:rad-prof} shows the
deprojected azimuthally averaged emission profile, assuming the same
inclination and position angle mentioned above. The emission is
centrally peaked and the surface brightness at the peak of the rings
decreases with radius. A striking aspect of the emission is the
difference between the inner 60~au disk, which consists of a central
component and three closely packed rings, and the outer $>60$~au disk
that consists of 2 bright rings that are well separated and spatially
resolved.  The two outer rings have been previously reported by
\citet{Fedele2018}. The new higher-angular resolution observations
reveal the inner disk is not smooth but contains substantial
substructure. In the inner disk the ring peaks are clearly resolved,
but they overlap at lower surface brightnesses. The gap near 100~au is
not completely empty of emission, as a faint ring can be seen in the
deprojected radial profile around this radius (better seen in the
zoomed panel in Fig.~\ref{fig:rad-prof}). An additional faint
component can also be seen at the edge of the disk, just outside the
bright ring located at $\sim120$~au.

In the next section, we model the emission in the {\it uv}-plane to obtain
the deconvolved position and width of the various rings observed
in the disk.

\subsubsection{Model-fitting in the {\it uv}-plane}

Given the striking ring-nature of the dust continuum emission, we
modeled the radial brightness distribution with the sum of concentric
Gaussian rings:
\begin{equation}
I(r) = \sum_{i=0}^N A_i \exp(-(r-r_i)^2/ 2\sigma_i^2 ).
\end{equation}
The number of rings is chosen through visual inspection of the cleaned
emission map and the deprojected radial profile (see
Figs.~\ref{fig:im} and \ref{fig:rad-prof}). We include a central
  component and 3 rings in the inner ($<60$~au) disk and 4 rings in
  the outer disk. The center position of the innermost Gaussian is
  fixed to zero, i.e., the disk center. Two of the outer rings
  correspond to the faint components near 100 and 130~au. We assume
  the emission is axisymmetric, and create synthetic visibilities
  given by the Hankel transform \citep{Pearson1999}:
\begin{equation}
V(\rho) = 2\pi \int^{\infty}_0 I_\nu(r) J_0(2\pi \rho r) r dr
\end{equation}
where $\rho$ is the deprojected {\it uv}-distance in units of
$k\lambda$, $r$ is the radial angular distance from the disk center in
units of radians, and $J_0$ is the zeroth-order Bessel function of the
first kind. 

\FigUVprof{}
\FigModel{}
\TabBestFitParamsT{}

To speed up the fitting method, the observed visibilities are first
deprojected, radially averaged and binned. We include {\it uv}-points
from 10 to 10000~k$\lambda$, in steps of 10~k$\lambda$. The
inclination ($i$), position angle (PA) and center of the disk, given
by an offset ($\delta_x$, $\delta_y$), are included as free parameters
in the fit. The total number of free parameters is then 27, that is 23
for the Gaussian rings ($A_i, r_i, \sigma_i$ with $i$ from 1 to 8;
$r_0=0$) and 4 for the disk geometry ($i$, PA, $\delta_x, \delta_y$).

We use the affine invariant MCMC sampler implemented in the \emcee{}
package \citep{Foreman-Mackey2013} to explore the parameter
space. Table~\ref{tab:best-fit} summarizes the resulting best-fit
parameters, calculated from the 50th percentile. The uncertainties are
calculated from the 16th and 84th percentiles.

We find a disk inclination of 34.9$^{\circ}$ and a position angle of
85.8$^{\circ}$, which are consistent with previous estimates
\citep{Fedele2018}. These are also consistent with what
\citet{Huang2018a} finds by fitting rings directly to the cleaned
image. We also find a small offset from the phase center. We note that
this offset is a just a nuisance parameter, since the disk center has
been altered during the self-calibration process. The rings in the
inner disk are centered at 15, 27, 41~au, and have FWHM between 7 and
17~au. Given that the spatial resolution of the observations is 5~au,
the rings are all resolved. The two prominent rings in the outer disk
are located at 74 and 120~au. \citet{Fedele2018} found this second
ring to be located at 130~au instead of 120~au. The difference is due
to the presence of another much fainter ring or bump, which we find to
be located at 140~au. The faint ring located in between the two
prominent outer rings is located at 92~au. We note that this ring is
not located at the gap center but is instead closer to the inner ring
near 74~au. These two fainter rings in the outer disk are found to be
much broader (FWHM of 23~au) than the rest of the rings, in particular
the two prominent outer rings which have FWHM of $\sim7-10$~au. This
suggest that the nature of these components may be different from the
rest, and their distribution are thus not well represented by a
Gaussian. They could instead be treated as faint emission that is
somehow connected to the brighter neighbor rings.

Fig.~\ref{fig:uv-prof} shows the observed real part of the deprojected
visibilities in black, using the best-fit disk inclination and
position angle. The imaginary part of the visibilities are shown in
light-blue (shifted by $-30$~mJy), and remain close to zero for all
{\it uv}-distance, consistent with symmetric emission of the disk. A
zoom of the shorter-baseline visibilities is shown in the upper-right
panel of the figure. Our simple parametric model of pure Gaussian
rings can recover most of the (very complicated) structure seen in the
radially averaged visibility profile. The cleaned image of the
best-fit model is shown in the left panel of
Fig.~\ref{fig:model-im}. A slightly hexagonal structure and bumpiness
can be seen in the rings, which is also seen in the observations. This
demonstrate these structures are not real but correspond to PSF
effects \citep[see][for a discussion]{Andrews2018}. The residuals,
corresponding to the cleaned image of residual visibilities
($V_{obs}-V_{model}$) are shown in the right panel of the figure. Our
best-fit model successfully reproduces the observations, as seen by
the low-level emission in the residuals map. However, some residuals
remain near the disk center. This suggests that some additional
substructure, in addition to Gaussian rings, are needed to fully
reproduce the inner disk structure.

%% \begin{itemize}
%% \item Describe the surface brightness radial profile (the emission
%%   consists of a series of concentric narrow rings; the emission is
%%   axysimmetric; high contrast for the outer rings).
%% \item Describe the fit done to the visibilities to extract the
%%   position and width of the ring components.
%% \item Describe the main results: 1) All the emission can be explained
%%   by ~7 concentric rings, 2) the rings are concentric, 3) a faint ring
%%   is found between the 2 outer rings, at around 100~au (the gap is not
%%   completely empty).
%% \item Compare results with Jane's fit to the image.
%% \end{itemize}

\subsection{$^{12}$CO $J=2-1$ emission}

\FigCO{} \FigCOchans{}

Figure~\ref{fig:co-im} shows the moment-zero map of the $^{12}$CO
$J=2-1$ line, integrated from $-3$ to 14~km~s$^{-1}$. Pixels with S/N
ratio lower than 3 have been clipped to highlight the substructure of
the emission. The West side of the disk is moderately affected by
cloud contamination due to the overlap between the velocities of the
cloud with the blue-shifted part of disk line emission. This explains
the East-West asymmetry. The bottom panel in Fig.~\ref{fig:co-im}
displays the azimuthally-averaged deprojected profile, including only
the East side of the disk. A selection of channel maps, showing the
emission from the East-side of the disk is shown in
Fig.~\ref{fig:co-chans}. Channel maps for the full velocity range can
be found in Fig.~\ref{fig:co-chans-all}. The $^{12}$CO $J=2-1$
emission 1) is centrally peaked, 2) extends much farther out than the
millimeter dust emission (out to 300~au), and 3) presents 4 main gaps
or emission decrements in the outer disk, near 45, 75, 120 and
210~au. We note this is not an effect of the clipping in the creation
the moment-zero map, as the gaps are also seen with even more clarity
in the individual channels (see Fig.~\ref{fig:co-chans}). The first
gap is located very close to the edge of the inner millimeter
disk. The next two gaps near 74 and 120~au spatially coincide with the
two prominent outer dust rings. This gap is better seen in the channel
maps, at a velocity of 5.8~km~s$^{-1}$. The spatial coincidence of
these two gaps with the location of the millimeter dust rings suggests
that the rings are optically thick and are absorbing the $^{12}$CO
emission (see the discussion in section~4.2). The fourth and outermost
gap seen in the $^{12}$CO emission is located at a distance of ~210~au
from the central star, well beyond the millimeter dust
edge. Therefore, it cannot be explained by dust opacity and must be a
real decrease in the $^{12}$CO column density at this radius.

The high-angular resolution observations of the $^{12}$CO $J=2-1$ line
allow us to determine the absolute orientation of the disk geometry
\citep{Rosenfeld2013}. Because the $^{12}$CO line is optically thick
and the disk is flared, the observed emission originates from the
surface layers of the disk which are elevated with respect to the
midplane, where the dust continuum emission arises. At high enough
angular resolution and if the disk is inclined enough, it is possible
to differentiate the front and back side of the disk. In the case of
AS~209, the southern part of the disk appears to be closer to us,
since the emission arising from the half-cone that is pointing towards
us is shifted to the north and appears brighter. By contrast, the
emission arising from the half-cone that is pointing to the back
appears dimmer and is shifted to the south (see
Fig.~\ref{fig:co-chans}). This effect is best seen at velocities of
5.8 and 6.15~km~s$^{-1}$. Compared to other disks, such as IM~Lup
\citep{Huang2018b} and HD~163296 \citep{Isella2018}, these effects are
subtle, however, which is likely due to the higher inclinations of
both disks compared to AS~209.

%% Observations of the scattered light emission from the AS~209 disk have
%% been recently presented by \citet{Avenhaus2018}. The emission is
%% surprisingly faint and compact compared to other disks of similar dust
%% mass and age. Moreover, AS~209 presents a large IR excess, which
%% usually results in bright scattered light emission. This could be the
%% result of self-shadowing of the inner disk, although the most likely
%% alternative is that the disk is very settled which hampers the
%% efficient scattering of photons from the star.

Scattered light observations trace the illumination of the small
grains in the disk. It is useful to compare scattered light to
$^{12}$CO observations since they both trace the upper layers of the
disk. The scattered light image of AS~209 presented recently by
\citet{Avenhaus2018} was found to be very faint and featureless in
comparison to other disks of similar mass. Although the scattered
light is faint, it is detected out to a radius of 200~au. The faint
nature of the scattered light could be the result of shadowing of the
outer disk by one of the rings in the inner disk, most likely the one
observed in millimeter continuum near 15~au. Another possibility is
that the disk has experienced substantial settling and it is very
flat, which hampers the efficient scattering of photons from the
star. From these observations, the authors computed the deprojected
profiles for two possible scenarios in which the North and the South
part of the disk are closer to us. As explained above, the new
$^{12}$CO observations demonstrate the latter alternative is the
correct one. For this scenario, the resulting radial profile of the
scattered light emission presents three rings, two of them located
near the two prominent outer dust rings seen by ALMA, and a third one
near 250~au. They also find a gap near 200~au. This outer ring and gap
coincide with the gap and outer ring seen in the $^{12}$CO emission.

%\item Dust gaps are not empty. CO emission is present within the gaps.
%\item CO emission extends outside the millimeter dust edge. This shows the
%  radial drift of large dust grains. Put limit on dust/co emission?

% Note the change of the slope at ~50 au.

  %\item observed surface brightness contrast, can we say something about
%  the lifetime of the rings in these pressure bumps?
%\item Dust opacity: where does the disk become optically thin/thick?

\section{Discussion}
\label{sec:discussion}

Although other disks are known to harbor multiple rings in their
millimeter emission, the disk around AS~209 is unique because of the
striking difference between the substructure in the inner and outer
disk, and the symmetric nature of the rings. While the inner disk
consists of closely-packed rings that resemble the emission seen in
TW~Hya and HD~163296, the outer disk harbors two prominent rings that
are separated by a gap with a much higher contrast compared to the
other disks. In this section we discuss the different possible origins
for these substructures, in both the dust continuum and the $^{12}$CO
line emission.

\subsection{Origin of the dust emission ring-morphology}

One of the main results from the DSHARP Large Program is that rings
and gaps are very common in protoplanetary disks \citep{Huang2018a},
but the origin of these substructures is still unknown. Several
hypotheses have been proposed, the most popular being the presence of
planets. An embedded planet can induce a gap opening (or multiple
gaps) by dynamic interactions with the disk. The depth of the gap will
depend on several factors, including the mass of the planet, the time
the planet has had to carve the gap, the disk aspect ratio $h/r$, and
the disk viscosity -- planets will open deeper gaps in low viscosity
disks \citep[e.g.][]{Crida2006,Dong2017,Bae2017}. Using 3D
hydro-dynamical simulations, \citet{Fedele2018} found that the
position, width and depth of the outermost gap at 100~au in the AS~209
disk is consistent with the presence of a 0.2~M$_J$ planet located at
95~au. Their simulation also predicted the presence of a feature
inside the gap. The feature was not detected in their observations but
it is now detected with the DSHARP observations. \citet{Fedele2018}
also found that the same planet could produce the other prominent gap
in the continuum at 60~au. Alternatively, \citet{Dong2018} showed
  that a $\sim0.1$~M$_J$ planet located at 80~au can produce the two
  major gaps seen in the outer AS~209 disk (at 60 and 100~au), and
  predicted an additional gap near 40~au, which is now detected by
  DSHARP. Another possibility, also consistent with the observations,
is the presence of a second $\sim0.1$~M$_J$ planet in the inner gap at
57~au, close to a 2:1 resonance with the outer planet
\citep{Fedele2018}.

The higher-angular resolution observations reveal the presence of
additional gaps in the inner disk that were unresolved in the
$\sim0\farcs17$ resolution observations of
\citet{Fedele2018}. Although they did not resolve the three individual
rings in the inner disk, they did report a kink in the profile near
$20-30$~au. The multiple rings seen in the inner disk could be
produced by planets in the outer disk. Hydrodynamics simulations have
shown that a single super-Earth planet could produce major gaps both
interior of the planet location and in the outer disk
\citep[e.g.,][]{Bae2017,Dong2017b}. A particular new feature in the
AS~209 disk is the gap located at roughly 10~au. The gap is not
resolved and it is not seen in the scattered light image, perhaps
because it is hidden by the $0\farcs185$ coronagraph
\citep{Avenhaus2018}. Inspired by the disk structures observed in the
DSHARP sources, \citet{Zhang2018} presented a grid of hydrodymical
simulations. In particular for AS~209, they show that a single planet
located at 99~au could produce simultaneously the various rings seen
in the inner 60~au and outer disk in AS~209 if $\alpha$ varies
radially. The simulation is shown in Fig.~\ref{fig:simulation}. The
planet responsible for the gaps has a mass of 0.087~M$_J$. Quite
remarkably, this simulation is able to produce not only the large gap
near 100~au, but also matches the position of the gaps in the inner
disk, at 60~au, 35~au and even the one at 24~au. A detailed
description of the simulation as well as a synthetic image of the dust
continuum emission is given in \citet{Zhang2018}.

\FigSimulation{}

It is worth noting that the most prominent gaps seen in the dust
continuum are not seen as prominent features in the $^{12}$CO emission (but
see section~\ref{sec:discussion-co}). This does not rule out the planet
hypothesis, as theoretical simulations have shown that it is possible
for planets to open gaps in the dust while leaving the gas emission
relatively featureless
\citep{Paardekooper2004,Rosotti2016,Isella2016,Dipierro2017}, especially for
optically thick lines like $^{12}$CO. This is particularly true if the
planet has a low mass, which is the case for the putative planets in
the AS~209 disk \citep{Zhang2018}.

%% Numerical studies have
%%  shown that less massive planets do notopen a gap in the gas but
%%  effectively open a gap in the dust(Paardekooper & Mellema2004,2006;
%%  Picogna & Kley2015;Dipierro et al.2016; Rosotti et al.2016; Dipierro &
%%   Laibe2017).
 
%the single planet theory need a high contrast for the gap at 100~au.
%Our new observations are consistent with this... ?

%what is the gap width predicted by models? is this consistent with the observations?

%the first gat near 15~au is interesting. could it be due t a planet? it is still far away from the star though.. 

%% We dont see a clear gap in the CO emission. Numerical studies have
%% shown that less massive planets do notopen a gap in the gas but
%% effectively open a gap in the dust(Paardekooper & Mellema2004,2006;
%% Picogna & Kley2015;Dipierro et al.2016; Rosotti et al.2016; Dipierro &
%% Laibe2017).

It has also been suggested that the rings observed at millimeter
wavelengths could by produced by changes in the dust properties at the
location of snowlines of the main ices, like H$_2$O, NH$_3$, CO and
N$_2$ \citep{Zhang2015,Okuzumi2016}. In this scenario, material can be
concentrated at the location of condensation fronts, which is critical
for the formation of planetesimals. At these locations the
$\mu$m-sized and mm-sized dust particles would have grown to cm sizes
and larger, and would thus be invisible at millimeter wavelengths,
appearing as gaps in the observations. We can estimate the location of
these snowlines in the AS~209 disk, using the midplane temperature
derived by \citet{Andrews2009}. The two outer gaps in the disk, near
60 and 100~au, have temperatures of 20 and 15~K, which are comparable
to the condensation temperatures of pure $^{12}$CO and N$_2$ ices,
respectively \citep{Zhang2015}. The water snowline, which is supposed
to be the most efficient snowline to concentrate solids, is located
very close to the central star ($<2$~au), and is thus unresolved even
with the ALMA observations. It is worth remembering that gas
temperatures in disks are very uncertain. In particular, for the
AS~209 disk the gas temperature was derived by fitting a power-law to
lower angular resolution observations of the dust continuum
\citep{Andrews2009}. The location of snowlines could therefore be
shifted in the disk. A major problem with this scenario is that the
gaps in the AS~209 disk are very wide (almost 20~au for the outermost
gap) and also very depleted in dust. Although they may contribute some
in the formation of these gaps, it is hard to explain how condensation
fronts alone could produce such wide gaps. Moreover, dust coagulation
and disk evolution models that take into account the condensation and
evaporation of major volatiles do not find enhanced grain growth near
the $^{12}$CO snowline \citep{Stammler2017}. Only at the location of
the water snowline is dust growth found to be efficient enough to
produce strong features in the dust continuum emission
\citep{Birnstiel2010,Drcazkowska2017,Schoonenberg2017}.

Other alternatives mainly involve internal disk gas dynamics resulting
from the coupling between gas and magnetic field. These include zonal
flows due to magnetorotational instability (MRI) turbulence
\citep{Johansen2009,Simon2014,Bai2014}, ring formation at dead zone
boundaries \citep{Flock2015,Lyra2015}, or through spontaneous magnetic
flux concentration \citep{Bai2014,Bethune2017,Suriano2018}. The zonal
flow scenario generally produces radial density variations up to a few
tens of percent on scales of a few scale heights. While this may be
consistent with the rings found in TW Hya \citep{Andrews2016}, the
ring separation in AS~209 is much broader (e.g., 46~au separation for
disk scale height of $5-10$~au in the outer disk), and the contrast
between rings and gaps is also much larger. The dead zone scenario
relies on a specific location corresponding to a transition in the
radial resistivity profile well within 100~au, making it difficult to
reconcile with the widely-separated multi-ring structure seen in the
outer disk of AS~209. The last scenario offers large degrees of
freedom attributed to how magnetic flux threading the disks evolve,
though this process is very poorly understood, and existing studies
largely rely on simplified assumptions
\citep{Lubow1994,Okuzumi2014,Guilet2014,Bai2017}. Therefore, while
some scenarios are unlikely, better theoretical understandings are
needed to assess the possibility of magnetic origin for the
substructures in AS~209.

%% Various mechanisms have been proposed in the literature thatcan be
%% assigned to three main categories: structures caused byfluid dynamics,
%% dust evolution effects, and planet–diskperturbations. More precisely,
%% these possibilities include zonalflows from magneto-rotational
%% instability(e.g., Simon &Armitage2014; Béthune et al.2016), gap/bump
%% structures inthe surface density close to the dead-zone outer
%% edge(e.g.,Flock et al.2015; Pinilla et al.2016; Ruge et
%% al.2016),efficientparticle growth at condensation fronts near ice
%% lines or adepletion of solid material between ice lines(Zhang et
%% al.2015;Pinilla et al.2017; Stammler et al.2017), aggregate
%% sinteringzones(Okuzumi et al.2016), secular gravitational
%% instabilities(Youdin2011; Takahashi & Inutsuka2014), or
%% planet–diskinteractions(e.g., Zhu et al.2011,2012; Donget
%% al.2015,2016; Rosotti et al.2016). Finally, dips or darkregions can be
%% interpreted as shadows by inner disk material(e.g., Marino et al.2015;
%% Pinilla et al.2015b; Stolkeret al.2016; Canovas et al.2017

%* are there any resonances between the rings and/or gaps?
  
\subsection{Origin of the $^{12}$CO $J=2-1$ emission rings and gaps}
\label{sec:discussion-co}

One way to produce gaps in the $^{12}$CO emission is by optically
thick dust emission. For a long time, dust opacity at millimeter
wavelengths was thought to be negligible, at least in the outer
disk. However, recent low-angular observations of a large sample of
disks \citep{Tripathi2017} as well as high-angular resolution
observations cast doubt on this assumption \citep[see also the case of
  HD~163296;][]{Isella2018}. In the case of AS~209, the peak
brightness temperature of the two prominent outer rings is
$\sim0.15-0.20$~mJy, which correspond to Rayleigh-Jeans brightness
temperatures of $\sim3-4$~K, and would only produce an optical depth
of $\sim0.4-0.5$ even assuming relatively conservative dust
temperatures of $\sim12-15$~K \citep[see also][]{Dullemond2018}.  The
dust rings therefore seem to be optically thin. One possibility to
explain the decrease of $^{12}$CO emission close to the location of
these two dust rings is that the rings are not resolved and are clumpy
in nature. A filling factor of $1/3$ would be enough to reproduce the
ratio between the dust temperature and the brightness temperature of
the rings. Another possible explanation is that some emission was
removed during the continuum subtraction process. Indeed, part of the
continuum can be absorbed by the molecule at the line center,
especially when the optical depth of the line is much higher than the
optical depth of the continuum. This can lead to an overestimation of
the dust contribution at the line center, which is estimated from the
line free channels, resulting in an underestimation of the line
emission \citep[see][]{Boehler2017,Weaver2018}.

\citet{Huang2016} presented CO isotopologue observations of the AS~209
disk at 0\farcs6 angular resolution, and found that the C$^{18}$O
emission consists of a central peak and a ring component centered at a
radius of $\sim$150~au. This ring component, however, was not observed
in their optically thick $^{12}$CO emission. The new higher angular
resolution $^{12}$CO observations reveal some new interesting
substructure. In particular, we detect a fainter outer ring centered
at $\sim240$~au, which was not detected in the C$^{18}$O emission
probably due to the lower signal-to-noise of the emission in the outer
disk. The CO distribution thus appears to have at least 2 ring
components located outside the millimeter dust disk and also well
outside the expected location of the CO snowline \citep[between 30 and
  90~au;][]{Huang2016}. \citet{Huang2016} suggested that the C$^{18}$O
ring at 150~au is caused by CO being desorbed back into the gas-phase,
which could happen by some non-thermal process (cosmic-rays or
high-energy photons), or by thermal inversion due to dust migration
\citep[e.g.,][]{Cleeves2016}. However, the presence of a second ring
at 240~au casts doubt on this interpretation. If chemistry is the
cause of the outermost $^{12}$CO ring, then the inner ring should be
produced by a different mechanism. Another alternative is that the
outermost gap near 210~au corresponds to a reduction of the total gas
density at this location. The origin of this gap is unclear, but we
can speculate it is produced by a planet. The formation of planets
this far from the star is hard to explain theoretically, but indirect
evidence of their existence has been found. \citet{Pinte2018} found
localized deviation from Keplerian velocities in the $^{12}$CO
emission, and attributed the observed velocity kink to a planet
located $\sim260$~au from the young HD~163296 star. We do not detect
any clear evidence of deviations from Keplerian velocities in the
AS~209 disk, but this could be due to lower signal-to-noise ratio and
spectral resolution compared to the HD~163296
observations. Alternatively, the velocity kink could be present in the
West side of the disk and thus hidden by the cloud contamination. We
do not detect any localized emission from a circumplanetary disk at
this radius in the dust continuum image either. Recently,
\citet{Teague2018} presented a new method to measure rotation curves
in disks. Using lower angular resolution archival CO data, they found
deviations of up to 5\% from Keplerian rotation at 250~au in the
AS~209 disk.

\section{Summary}
\label{sec:summary}

We have presented observations of the 1.2~mm dust continuum and the
$^{12}$CO $2-1$ line emission from the disk around the classical
AS~209 star, as part of the ALMA Large Program DSHARP. We have modeled
the dust emission in the {\it uv}-plane and find that the emission can
be well-represented by a series of narrow concentric rings. In
addition to the two prominent rings located at 74 and 120~au that were
previously reported by \citet{Fedele2018}, we find a central component
and three rings in the inner $<60$~au disk. Two main gaps are seen,
near 60 and 100~au. The second gap at 100~au is not completely empty
from dust grains, however, as we detect faint dust emission within the
gap. We also detect faint emission in the outer edge of the disk, out
to $\sim160$~au.

The $^{12}$CO image exhibits four gaps. Two of them spatially coincide
with the position of the two prominent outer dust rings, which suggest
the rings are optically thick. The outermost $^{12}$CO gap near
$\sim210$~au is located well-beyond the millimeter dust edge, and
therefore traces real $^{12}$CO depletion or a substantial decrease in
the gas density.

We discuss the different possibilities for the origin of the gaps seen
in the continuum emission. We find that, although some of the gaps
roughly coincide with the location of snowlines of major volatiles, it
it is unlikely that all the observed gaps are induced by a chemical
effect. This is because of the varied structures observed in the
different DSHARP sources, in terms of the width and contrast of the
gaps, and snowlines should produce similar configurations
\citep[see][for a discussion]{Huang2018a}. Recent hydro-dynamical
simulations show, however, that super-Earth planets can produce the
various rings seen in the continuum image, both in the inner disk and
in the outer disk. As the spatial resolution of the observations
improve, revealing new features, the locations and masses of these
planets can be better constrained. If the presence of a planet (or
multiple planets) in the AS~209 disk is confirmed, then planet
formation starts very early (few Myr) in the evolution of
disks. Moreover, if a planet is responsible for the gap seen in CO at
$\sim210$~au, then planets can form at large distance from the central
star, which challenges our current understanding of the planet
formation process.\\

\begin{acknowledgements}
V.V.G. and J.C acknowledge support from the National Aeronautics and
Space Administration under grant No. 15XRP15\_20140 issued through the
Exoplanets Research Program. S. A. and J. H. acknowledge funding
support from the National Aeronautics and Space Administration under
grant No. 17-XRP17\_2-0012 issued through the Exoplanets Research
Program. T.B. acknowledges funding from the European Research Council
(ERC) under the European Union’s Horizon 2020 research and innovation
programme under grant agreement No 714769. J.H. acknowledges support
from the National Science Foundation Graduate Research Fellowship
under Grant No. DGE-1144152. L. R. acknowledges support from the ngVLA
Community Studies program, coordinated by the National Radio Astronomy
Observatory, which is a facility of the National Science Foundation
operated under cooperative agreement by Associated Universities,
Inc. Z. Z. and S. Z.acknowledges support from the National Aeronautics
and Space Administration through the Astrophysics Theory Program with
Grant No. NNX17AK40G and Sloan Research Fellowship. Simulations are
carried out with the support from the Texas Advanced Computing Center
(TACC) at The University of Texas at Austin through XSEDE grant TG-
AST130002. C.P.D. acknowledges support by the German Science
Foundation (DFG) Research Unit FOR 2634, grants DU 414/22-1 and DU
414/23-1. M.B. acknowledges funding from ANR of France under contract
number ANR-16-CE31-0013 (Planet Forming disks). L.P. acknowledges
support from CONICYT project Basal AFB-17002 and from FCFM/U. de Chile
Fondo de Instalaci\'on Acad\'emica. This paper makes use of the
following ALMA data: ADS/JAO.ALMA\# 2016.1.00484.L, ADS/JAO.ALMA\#
2013.1.00226 and ADS/JAO.ALMA\# 2015.1.00486.S.  ALMA is a partnership
of ESO (representing its member states), NSF (USA) and NINS (Japan),
together with NRC (Canada), NSC andASIAA (Taiwan), and KASI (Republic
of Korea), in cooperation with the Republic of Chile. The Joint ALMA
Observatory is operated by ESO, AUI/NRAO and NAOJ.
\end{acknowledgements}

\appendix

\FigCOchansALL{}
  
%\FigCOpanels{}

%\FigWalkers{}
%\FigTriangle{}

\end{document}